\documentclass[conference]{IEEEtran}
\IEEEoverridecommandlockouts
\usepackage{cite}
\usepackage{amsmath,amssymb,amsfonts}
\usepackage{algorithmic}
\usepackage{graphicx}
\usepackage{textcomp}
\usepackage{xcolor}
\def\BibTeX{{\rm B\kern-.05em{\sc i\kern-.025em b}\kern-.08em
    T\kern-.1667em\lower.7ex\hbox{E}\kern-.125emX}}

\usepackage{authblk}
\usepackage{algorithm}

\usepackage[colorlinks=true, linkcolor=blue, citecolor=blue, urlcolor=blue]{hyperref}
\def\mrm{\mathrm}
\def\mbf{\mathbf}
\def\beq{\begin{equation}}
\def \eeq{\end{equation}}
\def\bbmat{\begin{bmatrix}}
\def\ebmat{\end{bmatrix}}

\def\E{\mathbb{E}}
\def\R{{\mathbb{R}}}
\def\C{{\mathbb{C}}}

\def\cn{\mathcal{CN}}

\def\jcs{JC\&S}
\def\Nsc{N_{sc}}
\def\Nap{N_{ap}}
\def\Nue{N_{ue}}
\def\setA{\mathcal{A}}
\def\setAt{\mathcal{A}_t}
\def\setU{\mathcal{U}}
\def\setS{\mathcal{S}}
\def\setD{\mathcal{D}}

\def\pmax{P_\text{max}}

\def\metric{\mathcal{M}}

\begin{document}

\title{Coordinated Decentralized Resource Optimization for Cell-Free ISAC Systems
\thanks{This work was supported by the National Science Foundation (NSF) under Grant CCF-2322191.}
}

% Authors

\author{Mehdi Zafari}
\author{Rang Liu}
\author{A. Lee Swindlehurst}
\affil{Department of Electrical Engineering and Computer Science, University of California, Irvine}

\maketitle

% For CORDIS content add: CORDIS/

% Abstract

\begin{abstract}
Integrated Sensing and Communication (ISAC) is emerging as a key enabler for 6G wireless networks, allowing the joint use of spectrum and infrastructure for both communication and sensing.
While prior ISAC solutions have addressed resource optimization, including power allocation, beamforming, and waveform design, they often rely on centralized architectures with full network knowledge, limiting their scalability in distributed systems.
In this paper, we propose two coordinated decentralized optimization algorithms for beamforming and power allocation tailored to cell-free ISAC networks.
The first algorithm employs locally designed fixed beamformers at access points (APs), combined with a centralized power allocation scheme computed at a central server (CS).
The second algorithm jointly optimizes beamforming and power control through a fully decentralized consensus ADMM framework.
Both approaches rely on local information at APs and limited coordination with the CS.
Simulation results obtained using our proposed Python-based simulation framework evaluate their fronthaul overhead and system-level performance, demonstrating their practicality for scalable ISAC deployment in decentralized, cell-free architectures.
\end{abstract}


%-------------------------
% Section: Introduction
%-------------------------

\section{Introduction}
\label{sec:intro}

%paragraph
Integrated Sensing and Communication (ISAC) is envisioned as a cornerstone of 6G wireless systems, enabling joint utilization of spectrum, hardware, and signal processing resources for communication and sensing~\cite{liu2022dual-functional}.
% By embedding sensing functionality into communication infrastructure, ISAC promises more efficient network operation and support for emerging applications such as real-time environmental awareness, autonomous vehicles, smart environments, and industrial automation.
Recent research has made significant strides in optimizing various ISAC functionalities, including power allocation (PA)~\cite{nguyen2025powerallocation, li2025sensingorientedadaptiveresourceallocation}, beamforming (BF)~\cite{rang2025beamforming}, waveform design~\cite{mura2025waveformdesign}, and distributed computation frameworks~\cite{lou2025BFopt, xiangnan2025federatedISAC}.
Despite progress, most existing ISAC solutions rely on centralized architectures, where a cloud central server (CS) collects global channel state information (CSI) and performs joint optimization across the network.
While such methods may theoretically yield performance gains, in practice, they suffer from critical limitations.
Centralized schemes often face bottlenecks due to scalability issues, excessive fronthaul communication, and latency, making them ill-suited for real-time or large-scale deployments.
More importantly, these approaches lack adaptability to distributed environments, where local processing and decentralized decision-making are essential due to physical or architectural constraints.

%paragraph
To address these challenges, we develop coordinated decentralized algorithms for BF and PA in cell-free ISAC systems, with the goal of distributing the computation across the network.
The first algorithm employs locally designed fixed beamformers at access points (APs) based on predefined methods, while solving a global PA problem centrally at the CS with limited network-level coordination.
This design is computationally efficient and scalable, but it restricts the number of simultaneously served users at each AP to its number of antennas (RF chains).
To overcome this limitation, we propose a second algorithm that jointly optimizes BF and PA through a fully decentralized consensus ADMM framework.
Both algorithms are designed to operate with local CSI and minimal information exchange over the fronthaul network, offering scalable solutions for distributed ISAC deployments.

%paragraph
In contrast to existing work that either assumes global CSI or leverages hierarchical and federated learning schemes with limited scalability~\cite{lou2025BFopt, xiangnan2025federatedISAC, zou2024distributedsensing}, our approach emphasizes simple coordination and tractable optimization in a fully decentralized setting.
The main contributions are summarized as follows:
\begin{itemize}
    \item \textbf{SplitOpt Algorithm}: We propose an algorithm that employs fixed local BF design at the APs and solves a global PA problem at the CS with limited coordination overhead.
    \item \textbf{JointOpt Algorithm}: We propose a second algorithm that jointly optimizes BF and PA through a fully decentralized consensus ADMM approach, enabling scalability and robustness in locally low-rank scenarios.
    \item \textbf{Open-Source Simulation Framework}: We provide a publicly available Python-based simulation framework that implements the proposed algorithms and supports reproducible and extensible research in distributed ISAC.
    \item \textbf{Comprehensive Performance Evaluation}: %Through extensive simulations using our framework, we provide a detailed analysis comparing both algorithms in terms of sensing and communication performance, scalability, and fronthaul communication overhead. We also provide results for centralized solution of the joint design problem, and compare the results.
    Using our simulation framework, we conduct extensive evaluations of both proposed algorithms, analyzing their sensing and communication performance, scalability, and fronthaul communication overhead. In addition, we implement a centralized solution for the joint BF and PA optimization problem and compare its performance against our ADMM-based decentralized algorithm.
\end{itemize}
% Beyond the theoretical development, we provide a Python-based simulation framework that implements the proposed algorithms and is designed to facilitate further research and extensions in decentralized ISAC systems.
% Using this framework, we evaluate the two algorithms in terms of system performance and fronthaul overhead, offering insights into the trade-offs among algorithm complexity, network performance, and coordination cost.
Our results demonstrate the potential of the proposed decentralized optimization methods to serve as scalable solutions for resource allocation in cell-free ISAC systems.

%-------------------------
% Section: System Model
%-------------------------

\section{ISAC System Model}
\label{sec:sys-mod}

%paragraph
We consider an orthogonal frequency division multiplexing (OFDM) system with total operating bandwidth \(B\), number of subcarriers \(\Nsc\), subcarrier spacing \(f_{sc} = B / \Nsc\), and symbol period \(T_s = T_{sc} + T_{cp}\), where \(T_{sc} = 1 / f_{sc}\) and \(T_{cp}\) is the duration of the cyclic prefix. 
The network consists of a set \(\setA\) of APs connected to a CS, with $|\setA| = \Nap$, and
each AP in $\setA$ is equipped with an \(M\)-element antenna array. %array of antennas containing  number of antenna elements.
% , which has a steering vector $\mbf{a} (\phi, \theta)$ for an azimuth angle \(\phi\) and elevation angle \(\theta\).
% Each AP in $\setA$ is equipped with a uniform circular array (UCA) of antennas with \(M\) number of antenna elements and UCA radius of $r$.
% The steering vector of the UCA for an azimuth angle \(\phi\) and elevation angle \(\theta\) is given as
% \begin{equation}
%     \mbf{a} (\phi, \theta) % = [e^{j\frac{2\pi r}{\lambda} \sin(\theta) \cos(\phi - \phi_1)}, \cdots, e^{j\frac{2\pi r}{\lambda} \sin(\theta) \cos(\phi - \phi_{M})}]^\top,
%     % = \{e^{j\frac{2\pi r}{\lambda} \sin(\theta) \cos(\phi - \phi_m)}\}, \; m=1, ..., M,
%     = \{e^{j 2\pi \hat{r} \sin(\theta) \cos(\phi - \phi_m)}\}, \; m=1, ..., M,
% \end{equation}
% for \(\phi_m = 2\pi (m-1) / M\) being the angular position of the $m$-th element around the circle, and $\hat{r}$ being the ratio of the UCA radius $r$ to the wavelength. % $\lambda$ being the wavelength.
A set $\setU$ of $\Nue$ single-antenna communication users (UEs) is present in the downlink, and one data stream is allocated per user. The set $\setS$ indexes the downlink sensing streams with $|\setS| = N_s$.
The complete set of downlink streams sharing the same time-frequency resources is indexed by the set $\setD = \setU \cup \setS$.
We consider a resource block (RB) consisting of $Q \leq \Nsc$ subcarriers and $K$ OFDM symbols.
% In our framework, we assume that the CS is responsible for scheduling RBs to serve both communication UEs and sensing tasks (e.g., detection and tracking).
% Depending on the scheduling strategy, the CS may allocate time-frequency resources for communication and sensing tasks with RB-orthogonal (RB-O) scheduling, i.e., orthogonal (non-overlapping) RBs are allocated separately to communication and sensing, or with RB-joint (RB-J) scheduling, i.e., selected RBs are shared for joint communication and sensing (\jcs) purposes.
% Alternative scheduling strategies may also be considered, where RBs for communication and sensing have partial overlap.
% AP clustering, user grouping (i.e., assignment of users to shared RBs), target scheduling (i.e., selection of sensing tasks or target locations), and RB scheduling (RB-O or RB-J) are centrally carried out at the CS, leveraging network state information including topology, service-level priorities, target coordinates, and AP availability for transmission and reception.
We assume that AP clustering and RB scheduling decisions are performed at the CS and broadcast to all APs, and that the APs are clustered into transmit and receive subsets, $\setA_t$ and $\setA_r$, to support multi-static sensing.
In this scenario, we focus specifically on PA and BF optimization for the case where a single RB is allocated to joint communication and sensing (\jcs).
To this end, we develop decentralized algorithms that reduce fronthaul communication overhead between the APs and the CS.

%-------------------------
% Sub-section: Comm Model
%-------------------------
\subsection{Communication Model}

%paragraph
The channel between UE $u$ and AP $a$ on a given subcarrier and OFDM symbol (resource element) in the selected RB is 
%modeled using a 3GPP-like Rician channel~\cite{overview2021, 3gpp_R17} for outdoor urban environments, 
denoted by $\mbf{h}_{au}$.
% \begin{align}
%     \mbf{h}_{qk,au} (t) = \beta_{au} e^{-j2\pi (\tau_{au} + \tau_o(t)) q f_{sc}} e^{j2\pi (f_{D,au} + f_o(t)) k T_s} \nonumber \\
%     \left(\sqrt{\frac{\riceK}{\riceK + 1}} \mbf{a} (\phi_{au}, \theta_{au}) + \sqrt{\frac{1}{\riceK + 1}} \widetilde{\mbf{h}}\right),
% \end{align}
% in which, $\beta_{au}$ represents the path-loss between UE $u$ and AP $a$, $\tau_{au}$ is the path delay, $\phi_{au}$ and $\theta_{au}$ are 3D angles of departure (AoD), $\riceK$ is the distance-dependent Rician K-Factor, $\widetilde{\mbf{h}} \sim \cn (0, \mbf{I})$ is the non-line-of-sight (NLoS) Rayleigh component, $f_{D,au}$ is the Doppler frequency, and $\tau_o(t)$ and $f_o(t)$ are time-variant timing offset (TO) and carrier frequency offset (CFO) due to imperfect synchronization.
We adopt a block-fading channel model where the given random channel realization is constant over the duration of the \jcs\ RB.
Moreover, each AP is assumed to have perfect CSI for all UEs on the \jcs\ RB.
The subsequent analysis is conducted on a per-subcarrier basis; hence, for notational simplicity, we omit the subcarrier index when it is clear from the context.

%paragraph
The downlink signal at AP $a$ for the $k$-th symbol is
\begin{align}
    \mbf{x}_a[k] &= \sum_{i \in \setD} \mbf{w}_{ai} \mrm{s}_i[k] = \sum_{u \in \setU} \mbf{w}_{au} \mrm{s}_u[k] + \sum_{j \in \setS} \mbf{w}_{aj} \mrm{s}_j[k] \nonumber \\
    &= \mbf{W}_a \mbf{s} [k] = \mbf{W}_a^{(c)} \mbf{s}^{(c)} [k] + \mbf{W}_a^{(s)} \mbf{s}^{(s)} [k],
\end{align}
where $\mbf{w}_{ai} \in \C^{M}$ is the BF vector for the $i$-th stream and $\mrm{s}_i[k]$ is the complex message on the $k$-th OFDM symbol.
The vectors $\mbf{w}_{ai}$ are collected as the columns of the BF matrix $\mbf{W}_a = [\begin{array}{c|c} \mbf{W}_a^{(c)} & \mbf{W}_a^{(s)} \end{array}] \in \C^{M\times (\Nue + N_s)}$, and we have $\mbf{s}[k] = \{\mrm{s}_i[k]\}, \forall i \in \setD$, where $\{\cdot\}$ represents the operator that collects the elements (or vectors) into a vector (or matrix) form.
The total available power at each AP is $\E[\|\mbf{x}_a[k]\|^2] = \pmax$, which can be allocated to both communication and sensing through the design of $\mbf{W}_a^{(c)}$ and $\mbf{W}_a^{(s)}$.
% We assume that the Doppler shift across OFDM symbols in one RB is small; thus $\mbf{h}_{k,au} = \mbf{h}_{au}, k = 1, \cdots, K$.
Therefore, the received signal at UE $u$ due to the $k$-th symbol is given as
\begin{align}
    \mrm{y}_u[k] % &= \sum_{a \in \setAt} \mbf{h}_{au}^H \mbf{x}_a[k] + \mrm{z}_u[k] \nonumber \\
    % &= \sum_{a \in \setAt} \sum_{i\in \setD} \mbf{h}_{au}^H \mbf{w}_{ai} \mrm{s}_i[k] + \mrm{z}_u[k] \nonumber \\
    &= \underbrace{\sum_{a \in \setAt} \mbf{h}_{au}^H \mbf{w}_{au} \mrm{s}_u[k]}_\text{Comm Desired Signal (CDS)} + \underbrace{\sum_{a \in \setAt} \sum_{u^\prime \in \setU \setminus u} \mbf{h}_{au}^H \mbf{w}_{au^\prime} \mrm{s}_{u^\prime}[k]}_\text{Multi-User Interference (MUI)} \nonumber \\
    & \quad + \underbrace{\sum_{a \in \setAt} \sum_{j \in \setS} \mbf{h}_{au}^H \mbf{w}_{aj} \mrm{s}_{j}[k]}_\text{Sensing-to-Comm Interference (S2CI)} + \underbrace{\mrm{z}_u[k]}_\text{Comm Noise (CN)},
\end{align}
where the noise term is assumed to be zero-mean with variance $\sigma_z^2$.
We define $\mbf{s}_i = [\mrm{s}_i[1], \cdots, \mrm{s}_i[K]]^\top \in \C^K, \forall i \in \setD$, as the vector of all $K$ samples from signal stream $i$,
% $\mbf{s}_i = \{\mrm{s}_i[k]\}, k\in [1,K], \forall i \in \setD.$
and we assume the radar and communication signals are statistically independent~\cite{liu2020jcr}, i.e., $\E[\mbf{s}_i \mbf{s}_i^H] = \mbf{I}_K$ and $\E[\mbf{s}_i \mbf{s}_j^H] = \mbf{0}_K$ for $i,j \in \setD$ and $i \neq j$.
Using these assumptions, the communication SINR for UE $u$ is given in \eqref{eq:comm_sinr}, presented at the top of the next page.

\begin{figure*}[!t]
% \hrulefill
\begin{equation}
    \label{eq:comm_sinr}
    \text{SINR}_u^{(c)} = \frac{\E[|\text{CDS}|^2]}{\E[|\text{MUI}|^2] + \E[|\text{S2CI}|^2] + \E[|\text{CN}|^2]} = \frac{|\sum_{a \in \setAt} \mbf{h}_{au}^H \mbf{w}_{au}|^2}{\sum_{u^\prime \in \setU \setminus u} |\sum_{a \in \setAt} \mbf{h}_{au}^H \mbf{w}_{au^\prime}|^2 + \sum_{j \in \setS} |\sum_{a \in \setAt} \mbf{h}_{au}^H \mbf{w}_{aj}|^2 + \sigma_z^2}
\end{equation}
\hrulefill
\end{figure*}

%paragraph
We further define the downlink channel matrix from AP $a$ to all users as $\mbf{H}_a = [\mbf{h}_{a1}, \cdots, \mbf{h}_{a\Nue}]^\top \in \C^{\Nue \times M}$.
We collect the noise-free signals received by all users in the vector
\begin{equation}
    \mbf{y}_\text{DL}[k] = \sum_{a\in \setAt} \mbf{H}_a \mbf{W}_a \mbf{s}[k] \triangleq \mbf{C}\, \mbf{s}[k] \in \C^{\Nue},
\end{equation}
for $\mbf{C} \triangleq \sum_{a\in \setAt} \mbf{H}_a \mbf{W}_a \in \C^{\Nue \times (\Nue + N_s)}$.
With a full-rank channel $\mbf{H}_a$ and an ideal BF matrix $\mbf{W}_a$, we can obtain matrix $\mbf{C} = \left[\begin{array}{c|c} \mbf{I}_{\Nue} & \mbf{0} \end{array}\right]$, i.e., the inter-user and sensing-induced interference is effectively canceled, so $\text{MUI} = \text{S2CI} = 0$.

%-------------------------
% Sub-section: Sensing Model
%-------------------------
\subsection{Sensing Model}

%paragraph
We assume that the CS dedicates a subset $\setA_r$ of the APs for receiving the sensing echo signals on the selected \jcs\ RB.
The sensing task considered in this work is single target detection through a multi-static setup.
We assume that the target location has a dominant line-of-sight (LoS) path to the APs in $\setAt$ and $\setA_r$.
For the sensing channel, we consider a single-point reflector with beam-space channel model~\cite{overview2021} 
% \begin{align}
%     \mbf{H}_{a_t a_r}^{(s)} [q,k] = \; &\beta_{a_t a_r} e^{-j2\pi \tau_{a_t a_r} q f_{sc}} e^{j2\pi f_{D,a_t a_r}kT_s} \nonumber \\
%     &\mbf{a}(\phi_{a_r}, \theta_{a_r}) \mbf{a}^H(\phi_{a_t}, \theta_{a_t}),
% \end{align}
\begin{align}
    \mbf{H}_{a_t a_r}^{(s)} [q,k] = \; &\beta_{a_t a_r} \Omega_{a_t a_r}[q,k] \mbf{a}(\phi_{a_r}, \theta_{a_r}) \mbf{a}^H(\phi_{a_t}, \theta_{a_t}),
\end{align}
in which, $\beta_{a_t a_r} \sim \cn(0, \zeta^2_{a_t a_r})$ is the channel gain that combines the path-loss and radar cross section (RCS) effect, $\Omega_{a_t a_r}[q,k] = \exp({-j2\pi \tau_{a_t a_r} q f_{sc}}) \exp({j2\pi f_{D,a_t a_r}kT_s})$ is the combined phase shift due to the path delay $\tau_{a_t a_r}$ and the Doppler frequency $f_{D,a_t a_r}$, $\phi_{a_r}$ and $\theta_{a_r}$ are the 3D angles of arrival (AoA) at AP $a_r$, and $\phi_{a_t}$ and $\theta_{a_t}$ are the AoDs from AP $a_t$.
The vector $\mbf{a} (\phi, \theta)$ represents the array steering vector for an azimuth angle \(\phi\) and elevation angle \(\theta\).
We use the Swerling-I model~\cite{book2010radar} for the RCS, allowing us to assume constant channel gain $\beta_{a_t a_r}$ across the entire \jcs\ RB.
The analysis is performed for each subcarrier; thus, the subcarrier index $q$ is dropped.
In the following, we use $\mbf{a}_{a_r}$ and $\mbf{a}_{a_t}$ to represent the array steering vectors at the receiver and transmitter, respectively, for notational brevity.
% , and the channel matrix can be written in a compact form as
% \begin{equation}
%     \mbf{H}_{a_t a_r}^{(s)} [k] = \beta_{a_t a_r} \Omega_{a_t a_r}[k] \mbf{a}_{a_r} \mbf{a}_{a_t}^H,
% \end{equation}
% in which, $\Omega_{a_t a_r}[k]$ models the phase shift components and $\mbf{a}_{a_r}$ and $\mbf{a}_{a_t}$ are the array steering vectors at the receiver and transmitter, respectively.

%paragraph
We also assume that the direct channels between transmit and receive APs are known to the receive AP, e.g., through pilot-based estimation prior to transmission on the \jcs\ RB.
Hence, the influence of the direct channels can be effectively removed.
The received signal for sensing at AP $a_r \in \setA_r$ at the $k$-th symbol time is % can be written as
\begin{align}
    \mbf{y}_{a_r}^{(s)} [k] % = &\sum_{a_t \in \setAt} \mbf{H}_{a_t a_r}^{(s)}[k] \mbf{x}_{a_t}[k] + \mbf{z}_{a_r}[k] \nonumber \\
    = &\underbrace{\sum_{a_t \in \setAt} \sum_{j \in \setS} \mbf{H}_{a_t a_r}^{(s)}[k] \mbf{w}_{a_t j} \mrm{s}_j[k]}_\text{Sensing Only (SO)} \nonumber \\
    &+ \underbrace{\sum_{a_t \in \setAt} \sum_{u \in \setU} \mbf{H}_{a_t a_r}^{(s)}[k] \mbf{w}_{a_t u} \mrm{s}_u[k]}_\text{Comm-to-Sensing Contribution (C2SC)} + \ \mbf{z}_{a_r}[k],
\end{align}
where both the SO and C2SC contribute to increasing the sensing parameter estimation accuracy.
By applying the receiver array steering vector $\mbf{a}_{a_r}$ as a matched filter to the received signal as $\mrm{r}_{a_r}[k] = \mbf{a}_{a_r}^H \mbf{y}_{a_r}^{(s)} [k]$, we obtain % $\mrm{r}_{a_r}[k] = M \sum_{a_t \in \setAt} \beta_{a_t a_r} \Omega_{a_t a_r}[k] \mbf{a}_{a_t}^H \mbf{W}_{a_t} \mbf{s}[k] + \widetilde{\mrm{z}}_{a_r}[k]$,
\begin{align}
    \mrm{r}_{a_r}[k] % &= \mbf{a}_{a_r}^H \mbf{y}_{a_r}^{(s)}[k] \nonumber \\
    % &= \sum_{a_t \in \setAt} \mbf{a}_{a_r}^H \beta_{a_t a_r} \Omega_{a_r a_r}[k] \mbf{a}_{a_r} \mbf{a}_{a_t}^H \mbf{x}_{a_t}[k] + \mbf{a}_{a_r}^H  \mbf{z}_{a_r}[k] \nonumber \\
    &= M \sum_{a_t \in \setAt} \beta_{a_t a_r} \Omega_{a_t a_r}[k] \mbf{a}_{a_t}^H \mbf{W}_{a_t} \mbf{s}[k] + \widetilde{\mrm{z}}_{a_r}[k],
\end{align}
where $\widetilde{\mrm{z}}_{a_r}[k] = \mbf{a}_{a_r}^H  \mbf{z}_{a_r}[k]$.
Collecting the filtered received signals over all $K$ symbols in the \jcs\ RB, we form the RB-level received signal at each AP $a_r$ as
\begin{align}
\label{eq:sense}
    \mrm{r}_{a_r}^\text{RB} %= \sum_{k=1}^K \mrm{r}_{a_r}[k] 
    = \underbrace{M \sum_{a_t \in \setAt} \beta_{a_t a_r} \mbf{a}_{a_t}^H \mbf{W}_{a_t} \sum_{k=1}^K \Omega_{a_t a_r}[k] \mbf{s}[k]}_\text{Sensing Desired Signal (SDS)} + \underbrace{\sum_{k=1}^K \widetilde{\mrm{z}}_{a_r}[k]}_\text{Sens Noise (SN)}.
\end{align}
% Using~\eqref{eq:sense} we can express the sensing SNR based on the filtered signal $\mrm{r}_{a_r}^\text{RB}$ at receive AP $a_r$ as the ratio of the SDS and sensing noise (SN) power:
Using~\eqref{eq:sense}, we can express the sensing SNR at receive AP $a_r$ based on the filtered signal \(\mrm{r}_{a_r}^\text{RB}\) as the ratio between the SDS power and the sensing noise (SN) power:
\begin{align}
    \label{eq:sensing_snr}
    \text{SNR}_{a_r}^{(s)} &= \frac{\E[|\text{SDS}|^2]}{\E[|\text{SN}|^2]} = \frac{M}{\sigma_{z,a_r}^2} \sum_{a_t\in\setAt} \zeta_{a_t a_r}^2 \|\mbf{a}_{a_t}^H \mbf{W}_{a_t}\|_2^2 \\
    &= \frac{M}{\sigma_{z,a_r}^2} \sum_{a_t} \zeta_{a_t a_r}^2 (\sum_{j\in \setS} |\mbf{a}_{a_t}^H \mbf{w}_{a_t j}|^2 + \sum_{u\in \setU} |\mbf{a}_{a_t}^H \mbf{w}_{a_t u}|^2), \nonumber
\end{align}
where $\sigma_{z,a_r}^2$ is the noise variance at AP $a_r$.
In the following section, we present optimization strategies aimed at maximizing communication SINR and sensing SNR metrics.

%-------------------------
% Section: Problem Formulation and Distributed Solutions
%-------------------------

\section{Problem Formulation and Distributed Algorithms} % CORDIS Distributed Optimization
\label{sec:dist-opt}

%paragraph
Our goal is to develop distributed optimization solutions for resource allocation that enable decentralized processing while minimizing information exchange over the fronthaul network.
In this section, we provide two optimization strategies for BF and PA.
The first approach splits the BF and PA optimization by allowing each AP to locally design its beamformers, while delegating PA to centralized optimization at the CS.
% This approach limits the maximum number of UEs simultaneously served by local beamformers to the number of antennas \textcolor{red}{(RF chains)} per AP.
This approach significantly reduces fronthaul overhead, but it limits the number of UEs that can be simultaneously served by the local beamformers at each AP to its number of antennas (RF chains).
Our second approach overcomes this limitation by formulating a multi-objective weighted sum maximization problem that jointly optimizes BF and PA.
To solve this problem in a fully decentralized manner, we develop an algorithm based on consensus ADMM.
These two approaches are detailed in the following subsections.

%-------------------------
% Sub-section: Split
%-------------------------
\subsection{SplitOpt (Distributed BF + Centralized PA)}

%paragraph
In the first approach, we adopt a fixed local BF method and optimize PA centrally at the CS.
To decouple BF and PA, we define a power splitting ratio (PSR) $\rho_a$ for each AP $a$, such that the communication power is given by $P_a^{(c)} = \rho_a \pmax$ and the sensing power by $P_a^{(s)} = (1-\rho_a) \pmax$.

\subsubsection{Distributed BF at APs}

%paragraph
For the local design of the beamformers, we will approximately solve the equations $\sum_{a\in \setAt} \mbf{H}_a \mbf{W}_a = \mbf{C}$ for the ideal $\mbf{C} = \left[\begin{array}{c|c} \mbf{I}_{\Nue} & \mbf{0} \end{array}\right]$ using regularized zero-forcing (RZF).
To decouple this problem for all APs $a \in \setAt$, we define the matrix $\mbf{C}_a = \alpha_a \mbf{C}$ and approximately solve the equation $\mbf{H}_a \mbf{W}_a = \mbf{C}_a$ for each $a \in \setAt$.
The weights are defined as $\alpha_a = \metric(a) / \sum_{a^\prime\in \setAt} \metric(a^\prime)$, where $\metric(a)$ is a metric designed to capture the quality of the channel matrix \(\mbf{H}_a\).
In this paper, we will set $\metric(a)$ as the inverse of the condition number of \(\mbf{H}_a\).
This choice implies that APs with poorly conditioned channels (i.e., higher condition numbers) will have a reduced contribution to the communication BF matrix.
To determine the weights $\alpha_a$, each AP sends its $\metric(a)$ value to the CS, and the CS broadcasts the sum $\sum_{a^\prime\in \setAt} \metric(a^\prime)$ to all transmit APs.
Then, each AP approximately solves the local system of equations $\mbf{H}_a \mbf{W}_a = \mbf{C}_a$ for $\mbf{W}_a$ using RZF, which leads to the solution
\begin{equation}
    \widetilde{\mbf{W}}_a^{(c)} = \left(\mbf{H}_a^H \mbf{H}_a + \varsigma \mbf{I}\right)^{-1} \mbf{H}_a^H \mbf{C}_a.
\end{equation}
After applying the power constraint, the BF matrix becomes
\begin{equation}
    \label{eq:lm-rzf}
    \mbf{W}_a^{(c)} = \sqrt{\rho_a \pmax} \ \widetilde{\mbf{W}}_a^{(c)} / \|\widetilde{\mbf{W}}_a^{(c)}\|_F,
\end{equation}
% where $\widetilde{\mbf{W}}_a^{(c)} = \left(\mbf{H}_a^H \mbf{H}_a + \varsigma \mbf{I}\right)^{-1} \mbf{H}_a^H \mbf{C}_a$.
which is designed at each AP for the communication users.
We refer to this BF approach as local modified RZF (LM-RZF).

%paragraph
For sensing, the transmit APs employ a null-space conjugate (NS-C) BF design~\cite{alkhateeb2024cf-isac} by projecting the beams pointed toward the target location onto the null space of the communication channel matrix.
The projection matrix at transmit AP $a$ is 
\begin{equation}
    \mbf{P}_a^{\perp} = \mbf{I} - \mbf{H}_a^{H} \left(\mbf{H}_a \mbf{H}_a^H + \epsilon \mbf{I}\right)^{-1} \mbf{H}_a,
\end{equation}
and by defining each sensing stream as $\widetilde{\mbf{w}}_{aj} = \mbf{P}_a^{\perp} \mbf{a}_{aj}$, we obtain the sensing BF matrix $\widetilde{\mbf{W}}_a^{(s)} = \{\widetilde{\mbf{w}}_{aj}\}$ for all $j\in \setS$.
By applying the power constraint, the NS-C BF matrix becomes
\begin{equation}
    \label{eq:ns-c}
    \mbf{W}_a^{(s)} = \sqrt{(1-\rho_a) \pmax} \ \widetilde{\mbf{W}}_a^{(s)} / \|\widetilde{\mbf{W}}_a^{(s)}\|_F,
\end{equation}
and the overall BF matrix $\mbf{W}_a$ is obtained by concatenating the communication and sensing beamformers.

\begin{algorithm}[t]
\caption{SplitOpt (DistBF-CentPA)}
\label{alg:splitOpt}
\begin{algorithmic}[1]

\STATE \textbf{Initialization:}
    \STATE \quad APs receive transmit signal information from the CS. % cluster formation, scheduling decisions, and

\STATE \textbf{Distributed BF at each AP $a\in\setAt$:}
    \STATE \quad Send $\metric(a)$ to CS and receive $\sum_{a^\prime} \metric(a^\prime)$ in return.
    \STATE \quad Employ LM-RZF using~\eqref{eq:lm-rzf} for communication UEs.
    \STATE \quad Employ NS-C using~\eqref{eq:ns-c} for target(s).
    \STATE \quad Measure $\widetilde{\alpha}_a$ and $\widetilde{\Delta}_a$ and share with the CS.

\STATE \textbf{Centralized PA at the CS:}
    \STATE \quad Receive vectors $\widetilde{\boldsymbol{\alpha}}$ and $\widetilde{\boldsymbol{\Delta}}$ from APs $a\in \setAt$.
    \STATE \quad Solve problem~\eqref{eq:opt_split} using projected gradient ascent.~\label{state:solve_opt}
    \STATE \quad Send optimal PSRs $\{\rho_{a}^*\}$ to APs $a\in \setAt$.
    
\end{algorithmic}
\end{algorithm}

\subsubsection{Centralized PA at CS}

%paragraph
To optimize the PA between communication and sensing, we define SINR-based and SNR-based utility functions for communication and sensing, respectively, and formulate a problem that maximizes sensing utility under a communication utility constraint. 
This formulation leads to the computation of optimal PSRs $\rho_a^*$ for each AP \(a \in \setAt\).
Assuming LM-RZF for communication BF design, which essentially removes MUI and S2CI interferences, the SINR~\eqref{eq:comm_sinr} reduces to approximately $(P_\text{max} / \sigma_z^2) \, U_\text{comm}$ for the communication utility function $U_\text{comm}$ defined as % $U_\text{comm} \triangleq \left|\sum_{a\in\setAt} \widetilde{\alpha}_a \sqrt{\rho_a}\right|^2 = (\widetilde{\boldsymbol{\alpha}}^\top \boldsymbol{\rho}^{1/2})^2$,
\begin{align}
    U_\text{comm} \triangleq \left|\sum_{a\in\setAt} \frac{\alpha_a}{\|\widetilde{\mbf{W}}_a^{(c)}\|_F} \sqrt{\rho_a}\right|^2 
    % \triangleq \left|\sum_{a\in\setAt} \widetilde{\alpha}_a \sqrt{\rho_a}\right|^2
    % = \left(\sum_{a\in\setAt} \widetilde{\alpha}_a \sqrt{\rho_a}\right)^2 
    = \left(\widetilde{\boldsymbol{\alpha}}^\top \boldsymbol{\rho}^{1/2}\right)^2,
\end{align}
in which $\widetilde{\alpha}_a = \alpha_a / \|\widetilde{\mbf{W}}_a^{(c)}\|_F$, $\widetilde{\boldsymbol{\alpha}} = \{\widetilde{\alpha}_a\}$, and $\boldsymbol{\rho} = \{\rho_a\}$, for all $a\in \setAt$.
Similarly, assuming NS-C for sensing BF design, the SNR-based sensing utility function is defined as % $U_\text{sens} \triangleq \sum_{a\in\setAt} \zeta_a^2 \left(\rho_a \|\mbf{a}_a^H \widehat{\mbf{W}}_a^{(c)}\|_2^2 - \rho_a \|\mbf{a}_a^H \widehat{\mbf{W}}_a^{(s)}\|_2^2\right) = \widetilde{\boldsymbol{\Delta}}^\top \boldsymbol{\rho}$,
\begin{align}
    U_\text{sens} &\triangleq \sum_{a\in\setAt} \zeta_a^2 \left(\rho_a \|\mbf{a}_a^H \widehat{\mbf{W}}_a^{(c)}\|_2^2 - \rho_a \|\mbf{a}_a^H \widehat{\mbf{W}}_a^{(s)}\|_2^2\right) \nonumber \\
    &= \sum_{a\in\setAt} \zeta_a^2 \rho_a \Delta_a \triangleq \widetilde{\boldsymbol{\Delta}}^\top \boldsymbol{\rho},
\end{align}
where $\widehat{\mbf{W}}_a = \widetilde{\mbf{W}}_a / \|\widetilde{\mbf{W}}_a\|_F$ for both communication and sensing matrices, $\Delta_a =  \|\mbf{a}_a^H \widehat{\mbf{W}}_a^{(c)}\|_2^2 - \|\mbf{a}_a^H \widehat{\mbf{W}}_a^{(s)}\|_2^2$, % the difference between the alignment of communication and sensing beamformers to the target direction,
$\widetilde{\Delta}_a = \zeta_a^2 \Delta_a$, and $\widetilde{\boldsymbol{\Delta}} = \{\widetilde{\Delta}_a\}$ for $a\in\setAt$.
Here, the subscript $a_r$ is removed for brevity and should be clear from the context.

%paragraph
The global PA optimization problem, which aims to maximize the sensing utility subject to a constraint on the communication utility, is formulated as follows:
\begin{subequations}
\label{eq:opt_split}
\begin{align}
    \max_{\{\rho_a\},\, \varepsilon \geq 0} \quad & \widetilde{\boldsymbol{\Delta}}^\top \boldsymbol{\rho} - \xi \varepsilon \label{eq:opt_split_obj} \\
    \text{s.t.} \quad & \widetilde{\boldsymbol{\alpha}}^\top \boldsymbol{\rho}^{1/2} + \varepsilon \geq \sqrt{\gamma}, \label{eq:opt_split_const1} \\
    \quad & 0 \leq \rho_a \leq 1, \ \forall a \in \setAt, \label{eq:opt_split_const2}
\end{align}
\end{subequations}
where $\gamma$ denotes the minimum communication utility requirement, $\varepsilon$ is a slack variable used to ensure the feasibility of constraint~\eqref{eq:opt_split_const1} for large values of $\gamma$, and $\xi$ is a parameter used to penalize the objective for slack usage.
% To balance sensing and communication needs, this trade-off parameter is adaptively tuned at the CS using service-level priorities, feedback from user experience, and sensing performance indicators.
% The objective function in \eqref{eq:opt_split_obj} is concave, and the constraints are convex.
The optimization problem in~\eqref{eq:opt_split} is a constrained concave maximization problem with convex constraints, which can be efficiently solved using the projected gradient ascent algorithm.
To solve this problem centrally at the CS, two real-valued vectors $\widetilde{\boldsymbol{\alpha}}$ and $\widetilde{\boldsymbol{\Delta}}$ must be shared with the CS.
Accordingly, each AP \(a \in \setAt\) sends its local values $\widetilde{\alpha}_a$ and $\widetilde{\Delta}_a$ to the CS and receives the optimal PSR $\rho_a^*$ in return.
The procedure for implementing the \textit{SplitOpt} algorithm is summarized in Algorithm~\ref{alg:splitOpt}.

%-------------------------
% Sub-section: DecentADMM
%-------------------------
\subsection{JointOpt (Decentralized ADMM)}

%paragraph
In the second approach, we formulate a joint BF and PA optimization problem and modify it to enable a decentralized solution using the ADMM algorithm~\cite{boyd2011book}.
For the optimization objective, we consider maximizing a weighted sum of utility functions for communication and sensing.
However, in this case, the communication utility is defined via a max-min formulation, aiming to maximize the minimum SINR across all UEs.
Hence, the communication utility is defined as % $U_\text{comm} \triangleq \min_{u\in\setU} \ \text{SINR}_u^{(c)}$,
\begin{equation}
    U_\text{comm} \triangleq \min_{u\in\setU} \ \text{SINR}_u^{(c)},
\end{equation}
for $\text{SINR}_u^{(c)}$ defined in~\eqref{eq:comm_sinr}.
We define the sensing utility function $U_\text{sens}$ using the SNR expression given in~\eqref{eq:sensing_snr} such that $\text{SNR}^{(s)} = (M/\sigma_z^2)\, U_\text{sens}$, which results in
\begin{equation}
    U_\text{sens} \triangleq \sum_{a\in\setAt} \zeta_{a}^2 \|\mbf{a}_{a}^H \mbf{W}_{a}\|_2^2.
\end{equation}
% The global optimization problem based on these utility functions will then be formulated as
% \begin{subequations}
% \label{eq:opt_admm_orig}
% \begin{align}
%     \max_{\mbf{W}_{a}} \quad & \lambda U_{\text{comm}} + (1 - \lambda) U_{\text{sens}} \label{eq:opt_admm_orig_obj} \\
%     \text{s.t.} \quad & \|\mbf{W}_a\|_F^2 % = \|\mbf{W}_a^{(c)}\|_F^2 + \|\mbf{W}_a^{(s)}\|_F^2
%     \leq \pmax, \ \forall a\in\setAt, \label{eq:opt_admm_orig_const}
% \end{align}
% \end{subequations}
The global optimization problem based on these utility functions will result in a non-smooth (due to the min operator) and non-convex (due to the SINR expression) objective function, making it challenging to apply standard distributed optimization methods without further reformulation.

%paragraph
To address these challenges, we employ an epigraph reformulation by introducing an SINR floor $\gamma$ that is maximized subject to the constraint that all UE SINRs are greater than this threshold.
Accordingly, the modified problem is given as
\begin{subequations}
\label{eq:opt_admm_mod}
\begin{align}
    \max_{\gamma, \mbf{W}_{a}} \quad & \lambda \gamma + (1 - \lambda) \sum_{a\in\setAt} \zeta_{a}^2 \|\mbf{a}_{a}^H \mbf{W}_{a}\|_2^2 \label{eq:opt_admm_mod_obj} \\
    \text{s.t.} \quad & \text{SINR}_u^{(c)} \geq \gamma, \ \forall u\in\setU, \label{eq:opt_admm_mode_const1} \\
    & \|\mbf{W}_a\|_F^2 \leq \pmax, \ \forall a\in\setAt, \label{eq:opt_admm_mode_const2}
\end{align}
\end{subequations}
where $\lambda \in [0, 1]$ is a trade-off parameter.
Larger values of $\lambda$ prioritize SINR for the UEs, whereas smaller values place greater emphasis on delivering power to the sensing direction.

Although problem~\eqref{eq:opt_admm_mod} features a linear objective in $\gamma$, the constraint~\eqref{eq:opt_admm_mode_const1} is still non-convex and coupled.
To decouple the problem, we adopt a local approximation of the SINR that can be independently evaluated at each AP $a\in\setAt$.
Following the similar SINR approximation in~\cite{zafari2024admm}, we use Minkowski's inequality to derive the local SINR contribution constraint as
\begin{equation}
    \label{eq:opt_admm_local_sinr_const}
    |\mbf{h}_{au}^H \mbf{w}_{au}|^2 \geq \eta_{au} \gamma_a (\psi_u + \sigma_z)^2, \ \forall u\in \setU,
\end{equation}
that forces each AP to contribute a fair share toward achieving the SINR target.
The parameter $\eta_{au}$ is a fractional weight that quantifies the responsibility of AP $a$ in contributing to the SINR of user $u$.
A straightforward choice is to assign equal responsibility to all APs by setting \(\eta_{au} = 1 / |\setAt|\).
However, alternative weighting strategies can also be adopted, e.g., based on channel gain or local channel quality (e.g., channel rank).
%, or network topology.
% This parameter can be uniformly chosen as \(\eta_{au} = 1 / \Nap\).

The interference produced by AP $a$ at user $u$ is proportional to $\psi_{au} = (\sum_{i\in\setD \setminus u} |\mbf{h}_{au}^H \mbf{w}_{ai}|^2)^{1/2}$, and we define $\psi_u = \sum_{a\in\setAt} \psi_{au}$ for each user $u$.
We further define the real-valued vector $\boldsymbol{\psi}=\{\psi_u\} \in \R^{\Nue}$ for all $u\in\setU$.
This interference vector must be known to all APs \(a \in \setAt\).
To handle the local constraint~\eqref{eq:opt_admm_local_sinr_const}, we adopt a standard inner-approximation (successive convex approximation) framework by replacing the non-DCP convex term $g(\mbf{w}_{au}) = |\mbf{h}_{au}^H \mbf{w}_{au}|^2$ with its first-order Taylor expansion around the previous iterate $\mbf{w}_{au}^{[t]}$ as
\begin{equation}
    \bar{g}(\mbf{w}_{au}; \mbf{w}_{au}^{[t]}) =  2 \Re \{(\mbf{h}_{au}^H \mbf{w}_{au}^{[t]})^* \, \mbf{h}_{au}^H \mbf{w}_{au}\} - |\mbf{h}_{au}^H \mbf{w}_{au}^{[t]}|^2,
\end{equation}
which yields an affine and DCP-compliant constraint and defines an inner approximation of the original feasible set.

% The resulting convex subproblem is implemented in CVX as a modeling layer, and solved numerically using standard convex solvers (e.g., SDPT3/SeDuMi).

% The local constraint~\eqref{eq:opt_admm_local_sinr_const} is still \textcolor{red}{not compatible with standard CVX solvers since it is  in the form of convex $\geq$ affine}.
% We adopt a linear approximation of $g(\mbf{w}_{au}) = |\mbf{h}_{au}^H \mbf{w}_{au}|^2$ based on the first-order Taylor approximation around a given point $\mbf{w}_{au}^{[t]}$ as % $\bar{g}(\mbf{w}_{au}; \mbf{w}_{au}^{[t]}) =  2 \Re \{(\mbf{h}_{au}^H \mbf{w}_{au}^{[t]})^* \mbf{h}_{au}^H \mbf{w}_{au}\} - |\mbf{h}_{au}^H \mbf{w}_{au}^{[t]}|^2$,
% \begin{equation}
%     \bar{g}(\mbf{w}_{au}; \mbf{w}_{au}^{[t]}) =  2 \Re \{(\mbf{h}_{au}^H \mbf{w}_{au}^{[t]})^* \, \mbf{h}_{au}^H \mbf{w}_{au}\} - |\mbf{h}_{au}^H \mbf{w}_{au}^{[t]}|^2,
% \end{equation}
% and we use this approximation instead \textcolor{red}{to make the problem solvable with standard CVX solvers.}

\begin{algorithm}[t]
\caption{JointOpt (DecentADMM)}
\label{alg:jointOpt_admm}
\begin{algorithmic}[1]

\STATE \textbf{Initialization:}
    \STATE \quad Initialize $\mbf{W}_a^{[0]}$, $\gamma_a^{[0]}$, and $\nu_a^{[0]}$ for all $a\in\setAt$.
    \STATE \quad CS broadcasts initial $\boldsymbol{\psi}^{[0]}$ to all APs $a\in\setAt$.
    % \STATE \quad Set iteration counter $t = 1$.

\STATE \textbf{Local Update at each AP $a\in\setAt$:}~\label{state:admm_start}
    \STATE \quad Obtain $\gamma^{[t-1]}$, $\boldsymbol{\psi}^{[t-1]}$, $\nu_a^{[t-1]}$, and $\xi^{[t-1]}$.
    \STATE \quad Solve the QCQP problem in \eqref{eq:opt_admm_final} via CVX.
    \STATE \quad Measure $\psi_{au}^{[t]}$ for all $u\in\setU$ and share with the CS.
\STATE \textbf{Global Update at the CS:}
    \STATE \quad SINR floor update $\gamma^{[t]} = \frac{1}{\Nap} \sum_{a\in\setAt} (\gamma_a^{[t]} + \nu_a^{[t-1]})$
    \STATE \quad Interference update $\psi_u^{[t]} = \sum_{a\in\setAt} \psi_{au}^{[t]}, \forall u\in\setU$.
    \STATE \quad Broadcast $\gamma^{[t]}$ and $\boldsymbol{\psi}^{[t]}$ to all APs $a\in\setAt$.
\STATE \textbf{Local Dual and Slack Updates:}
    \STATE \quad Update dual variable $\nu_a^{[t]} = \nu_a^{[t-1]} + (\gamma_a^{[t]} - \gamma^{[t]})$
    \STATE \quad Slack penalty update $\xi^{[t]} = \min(\xi_{\max}, \xi^{[t-1]}+\Psi t)$
% \IF{$\max_a |\gamma_a - \gamma| > \epsilon_\text{conv}$}
\STATE \textbf{Check Convergence:}
    \STATE \quad If $\max_a |\gamma_a - \gamma| < \epsilon_\text{conv}$, exit.
    \STATE \quad Else, set $t \leftarrow t+1$ and return to step \ref{state:admm_start}.

\end{algorithmic}
\end{algorithm}

%paragraph
To enable decentralized ADMM, we introduce a local consensus variable $\gamma_a$ for each AP \(a \in \setAt\), and enforce global consensus through a corresponding update step.
By employing the local SINR approximation and the consensus variable, the global problem in~\eqref{eq:opt_admm_mod} can be decomposed into parallel subproblems solvable independently at each AP \(a \in \setAt\).
The augmented Lagrangian for local updates is obtained as
\begin{align}
    \mathcal{L}_\varrho^{(a)} = \lambda \gamma_a + (1-\lambda) \|\mbf{a}_{a}^H \mbf{W}_{a}\|_2^2 - \frac{\varrho}{2} (\gamma_a - \gamma + \nu_a)^2,
\end{align}
where $\varrho$ is the penalty parameter, $\gamma_a$ and $\gamma$ are local and global consensus variables, respectively, and $\nu_a$ is the dual variable.
Since $f(\mbf{W}_{a}) = \|\mbf{a}_{a}^H \mbf{W}_{a}\|_2^2$ is not concave, the augmented Lagrangian  is nonconcave and cannot be handled directly.
We therefore adopt an inner-approximation (successive convex approximation) scheme and use
a concave surrogate for $f$ based on the first-order Taylor series expansion as %$\bar{f} (\mbf{W}_{a}; \mbf{W}_{a}^{[t]}) = 2 \Re\{(\mbf{a}_{a}^H \mbf{W}_{a}^{[t]}) \mbf{W}_{a}^H \mbf{a}_{a}\} - \|\mbf{a}_{a}^H \mbf{W}_{a}^{[t]}\|_2^2$
\begin{equation}
    \bar{f} (\mbf{W}_{a}; \mbf{W}_{a}^{[t]}) = 2 \Re\{(\mbf{a}_{a}^H \mbf{W}_{a}^{[t]}) \mbf{W}_{a}^H \mbf{a}_{a}\} - \|\mbf{a}_{a}^H \mbf{W}_{a}^{[t]}\|_2^2,
\end{equation}
which is concave and is used as a lower bound approximation.
The augmented Lagrangian defined by replacing $f(\mbf{W}_{a})$ with the concave surrogate $\bar{f} (\mbf{W}_{a}; \mbf{W}_{a}^{[t]})$ is denoted by $\bar{\mathcal{L}}_\varrho^{(a)}$.

%paragraph
To ensure that the local subproblems remain feasible, we introduce non-negative slack variables $\varepsilon_{au}$ for each user $u\in\setU$, and penalize them in the objective using a large weight $\xi$, which is updated every iteration.
% Using the slack variables serves to relax the SINR constraints, helping to prevent infeasibility—particularly during early ADMM iterations when the SINR target $\gamma$ may be overly ambitious.
Finally, the local optimization problem at each transmit AP $a\in\setAt$ has the following form
\begin{subequations}
\label{eq:opt_admm_final}
\begin{align}
    \max_{\gamma_a,\, \mbf{W}_{a},\, \varepsilon_{au}}  & \bar{\mathcal{L}}_\varrho^{(a)} - \xi \sum_{u\in\setU} \varepsilon_{au}~\label{eq:opt_admm_final_obf} \\
    \text{s.t.} \quad & \bar{g}(\mbf{w}_{au}; \mbf{w}_{au}^{[t]}) \geq \eta_{au} \gamma_a (\psi_u + \sigma_z)^2 - \varepsilon_{au},~\label{eq:opt_admm_final_const1} \\
    & \|\mbf{W}_a\|_F^2 \leq \pmax, \ \forall u\in\setU.~\label{eq:opt_admm_final_const2}
\end{align}
\end{subequations}
The local problem in \eqref{eq:opt_admm_final} is a quadratically-constrained quadratic program (QCQP), which can be efficiently solved using standard CVX solvers.
The complete procedure of the \textit{JointOpt} algorithm is summarized in Algorithm~\ref{alg:jointOpt_admm}.

%-------------------------
% Section: Simulation Results
%-------------------------

\section{Simulation Results}

%paragraph
In addition to the theoretical analysis, we developed a Python-based simulation framework, which is publicly available on GitHub at~\cite{zafari2025repo}.
The simulator implements the proposed algorithms and provides a flexible platform for further research and development on decentralized ISAC systems.
All datasets and seed numbers used for the figures in this paper are also provided in the repository.
We also provide a centralized solution for problem~\eqref{eq:opt_admm_mod}, with the full results made available in our repository. Due to space limitations, centralized results are included in the paper for only one simulation scenario.
We simulate an environment in which the APs are uniformly distributed on a circle of radius $650$m, and each is equipped with a uniform circular array (UCA) of size $M$.
The UEs and a single target are randomly placed within a concentric circle of radius $1$km, and the closest AP to the target location is selected as the sensing receive AP.
The ratio of $P_\text{max}$ to the noise power is $P_\text{max} / \sigma_z^2 = 20$ dB, the target RCS variance is $\zeta_a^2 = 0.5$ (-3 dB), and the communication channels are generated based on the 3GPP Rician model for urban micro cells~\cite{overview2021, 3gpp_R17}.

%Figures
\begin{figure}[!t]
    \centering
    \includegraphics[width=3.49in]{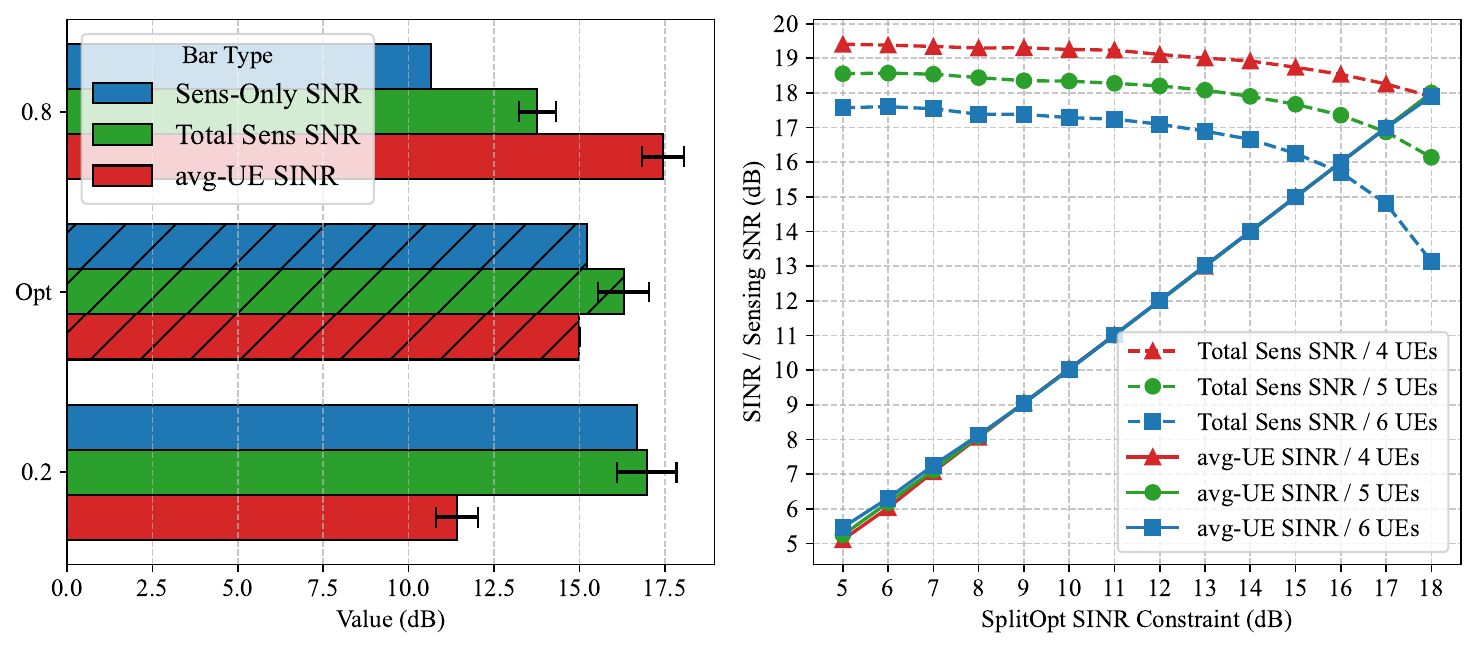}
    \caption{Performance of Algorithm~\ref{alg:splitOpt}. (Left) SINR and sensing SNR comparison for fixed PSRs (0.2, 0.8) and optimal PSRs (Opt). (Right) SINR and sensing SNR achieved under different SINR constraints $\gamma$ with \(\Nue \in \{4, 5, 6\}\).}
    \label{fig:splitOpt}
\end{figure}

%paragraph
To evaluate the performance of Algorithm~\ref{alg:splitOpt}, % we simulate a scenario with $\Nap = 4$ APs each with a UCA of size $M = 10$, $N_s = 1$ sensing target, and $\Nue = 6$ UEs.
we simulate a scenario with $4$ APs each with a UCA of size $10$, one sensing target, and $6$ UEs.
We compare the performance of local BF with fixed PSRs $\rho_a \in \{0.2, 0.8\}$ against the optimal PSRs $\rho_a^*$ obtained using the \textit{SplitOpt} algorithm.
The results are presented in Fig.~\ref{fig:splitOpt} (left), illustrating that a PSR of 0.2 prioritizes sensing performance, whereas a PSR of 0.8 favors higher communication SINR.
In contrast, the \textit{SplitOpt} algorithm satisfies an SINR constraint of 15~dB for all users while simultaneously achieving a high sensing SNR.
In Fig.~\ref{fig:splitOpt} (right), the SINR and sensing SNR results achieved by the \textit{SplitOpt} algorithm are presented for different utility constraints $\gamma$ and varying numbers of users.
The results show that the SINR constraint is satisfied across all cases, while the sensing SNR begins to decline once the SINR constraint exceeds approximately 13~dB.
Moreover, the degradation in sensing SNR is more pronounced as the number of users increases since satisfying the SINR constraint requires allocating a larger portion of the available power to communication.
This analysis demonstrates that Algorithm~\ref{alg:splitOpt} provides an effective decentralized solution, requiring minimal coordination with the CS.

%Figures
\begin{figure}[!t]
    \centering
    \includegraphics[width=3.49in]{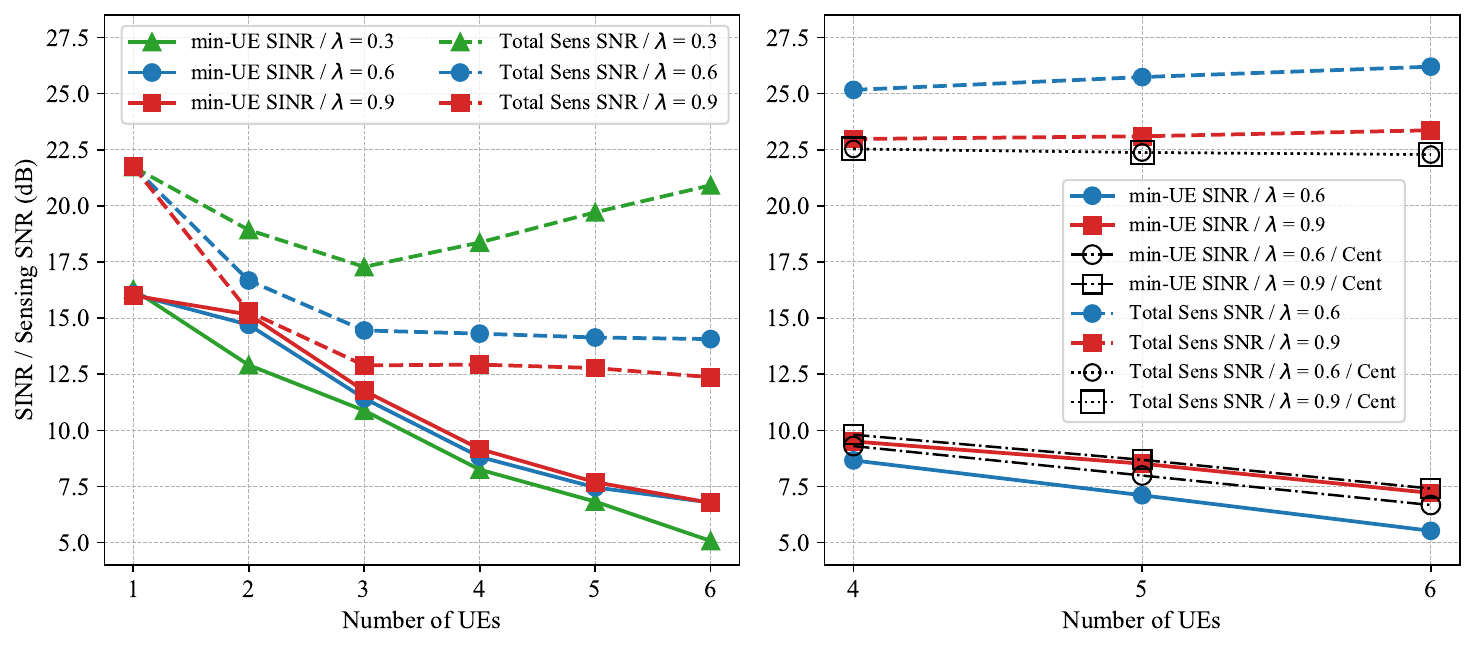}
    \caption{Performance of Algorithm~\ref{alg:jointOpt_admm}. (Left) SINR and sensing SNR vs different number of users for trade-off parameter $\lambda\in\{0.3,0.6,0.9\}$. (Right) SINR and sensing SNR with $\Nap = 11$ APs and $M=3$ antennas for both the proposed ADMM-based decentralized solution and the centralized solution.}
    \label{fig:jointOpt}
\end{figure}

%paragraph
Using the same simulation settings, except for varying the number of users from 1 to 6, we next used Algorithm~\ref{alg:jointOpt_admm} for joint BF and PA design.
The results are presented in Fig.~\ref{fig:jointOpt} (left) for the trade-off parameter values $\lambda \in \{0.3, 0.6, 0.9\}$.
The results show that as the number of users increases from 1 to 6, the minimum user SINR guaranteed by the \textit{JointOpt} algorithm decreases from approximately 16~dB to 6~dB.
Additionally, the advantage of reducing the trade-off parameter $\lambda$ to prioritize sensing performance is clearly illustrated: $\lambda = 0.3$ achieves the highest sensing SNR while maintaining only a slight reduction in SINR compared to higher $\lambda$ values. % communication (SINR)

\renewcommand{\arraystretch}{1.5}
\begin{table}[t]
\caption{Comparing Fronthaul Communication Overhead for each AP}
\label{tab:compare}
\centering
\begin{tabular}{|c|c|c|c|}
\hline 
 & \textbf{Centralized} & \textbf{SplitOpt} & \textbf{JointOpt} \\ \hline \hline
Data to share & $\mbf{H}_a$, $\mbf{W}_a$ & $\metric(a)$, $\widetilde{\alpha}_a$, $\widetilde{\Delta}_a$, $\rho_a^*$ & $\gamma^{[t]}$, $\boldsymbol{\psi}^{[t]}$ \\ \hline
Size of data & $2\times\C^{M\times |\setD|}$ & $4\times\R^{1}$ & $\R^1$, $\R^{|\setU|}$ \\ \hline
\# of scalars & $280$ & $4$ & $7\times T_\text{ADMM}$ \\ \hline
% Processors & $1$ & $\Nap$ & $\Nap$ \\ \hline
Scalable & NO & YES & YES \\ \hline
\end{tabular}
\end{table}

%paragraph
% Algorithm~\ref{alg:jointOpt_admm} is particularly advantageous in scenarios where the number of antennas per AP is smaller than the number of users, resulting in locally low-rank channels where Algorithm~\ref{alg:splitOpt} cannot be applied.
Algorithm~\ref{alg:jointOpt_admm} is particularly advantageous for scenarios with fewer antennas (RF chains) per AP than the number of users, where locally low-rank channels render Algorithm~\ref{alg:splitOpt} inapplicable.
To illustrate this, we simulate a setup with 11 APs (10 Tx APs) each with 3 antennas, serving \(\{4, 5, 6\}\) users and sensing a single target location.
% This scenario maintains the same total number of transmit antennas (30) as in the previous scenarios.
The total number of transmit antennas (30) is the same as in the previous scenarios.
% Fig.~\ref{fig:jointOpt} (right) shows that high sensing SNR is maintained while achieving a minimum user SINR between 5 and 10~dB.
% It also demonstrates that the central solution for problem~\eqref{eq:opt_admm_mod} achieves approximately the same performance as our decentralized algorithm.
Fig.~\ref{fig:jointOpt} (right) shows that high sensing SNR is maintained with a minimum user SINR between 5–10~dB, and that the centralized solution to problem~\eqref{eq:opt_admm_mod} performs comparably to our decentralized algorithm.
% These results confirm that both proposed algorithms offer efficient decentralized solutions for BF and PA design in cell-free ISAC systems, enabling flexible trade-offs suited to various network scenarios.
These results confirm that both algorithms enable efficient decentralized BF and PA in cell-free ISAC systems, supporting flexible trade-offs across diverse scenarios.
Table~\ref{tab:compare} compares the two algorithms against centralized solutions in terms of coordination overhead, highlighting a significant reduction in information exchange.
The number of real scalars to share over the fronthaul is provided for one of the scenarios with %4 APs, each with 10 antennas, 6 UEs, and one sensing target.
$M=10$ antennas per AP, serving $\Nue=6$ UEs and a single target.
$T_\text{ADMM}$ denotes the number of ADMM iterations, which was on the order of  5 to 10 for our simulations.

%-------------------------
% Section: Conclusion
%-------------------------

\section{Conclusion}

%paragraph
In this paper, we proposed two coordinated decentralized algorithms for BF and PA in cell-free ISAC systems.
The first algorithm splits BF and PA by employing fixed local beamformers and centralized power optimization with limited coordination.
The second algorithm jointly optimizes BF and PA through a fully decentralized consensus ADMM approach, enabling scalable operation even in the presence of locally low-rank channels.
% To support further research and reproducibility, we developed and publicly released a Python-based simulation framework implementing both algorithms.
To support further research and reproducibility, we developed and released a Python-based simulation framework implementing both algorithms.
Simulation results showed that, compared to centralized solutions, both algorithms achieve effective sensing and communication performance while significantly reducing fronthaul overhead.

% References
\bibliographystyle{IEEEtran}
\bibliography{ref}

\end{document}